\documentclass[aps,reprint,showpacs,showkeys,superscriptaddress]{revtex4-1}

\usepackage[pdftex]{color,graphicx}

\usepackage{amsfonts,amssymb,amsmath}
\usepackage[english]{babel}

\usepackage[separate-uncertainty=true]{siunitx}
\usepackage{hyperref}
\hypersetup{colorlinks=true,linkcolor=blue,citecolor=blue,filecolor=blue,urlcolor=blue}
\usepackage[sort&compress]{natbib}

\newif\ifusebibfile
\usebibfiletrue
\usebibfilefalse

\renewcommand{\textcolor}[2]{#2}
\newcommand{\figref}[2]{\hyperref[#1]{\ref{#1}(#2)}}
\newcommand{\ket}[1]{\mbox{$\vert #1 \rangle$}}

\renewcommand\textemdash{\leavevmode\unskip\kern0.8pt\rule[0.215\baselineskip]{8pt}{0.22pt}\kern1pt\ignorespaces}

\newcommand{\LGI}{LG inequality}

\begin{document}
\selectlanguage{english}

\makeatletter
\def\@pacs@name{Subject Areas: }%
\def\appendixname{APPENDIX}%
\makeatother

\title{Ideal negative measurements in quantum walks disprove theories based on classical trajectories}

\author{Carsten Robens}
\author{Wolfgang Alt}
\author{Dieter Meschede}
\affiliation{Institut f\"ur Angewandte Physik, Universit\"at Bonn,
Wegelerstr.~8, D-53115 Bonn, Germany}
\homepage{http://quantum-technologies.iap.uni-bonn.de}
\email{alberti@iap.uni-bonn.de}
\author{Clive Emary}
\affiliation{Department of Physics and Mathematics, University of Hull, Kingston-upon-Hull, HU6 7RX, United Kingdom}
\author{Andrea Alberti}
\affiliation{Institut f\"ur Angewandte Physik, Universit\"at Bonn,
Wegelerstr.~8, D-53115 Bonn, Germany}

\date{\today}

\pacs{
	Quantum Physics, Atomic and Molecular Physics
}


\begin{abstract}
We report on a stringent test of the non-classicality of the motion of a massive quantum particle, which propagates on a discrete lattice. Measuring temporal correlations of the position of single atoms performing a quantum walk, we observe a $6\,\sigma$ violation of the Leggett-Garg inequality. Our results rigorously excludes (\emph{i.e.}\ falsifies) any explanation of quantum transport based on classical, well-defined trajectories.
We use so-called ideal negative measurements \textemdash an essential requisite for any genuine Leggett-Garg test \textemdash to acquire information about the atom's position, yet avoiding any direct interaction with it.
The interaction-free measurement is based on a novel atom transport system, which allows us to directly probe the absence rather than the presence of atoms at a chosen lattice site.
Beyond the fundamental aspect of this test, we demonstrate the application of the Leggett-Garg correlation function as a witness of quantum superposition. We here employ the witness to discriminate different types of walks spanning from merely classical to wholly quantum dynamics.
\end{abstract}

\maketitle

The superposition principle is one of the pillars of quantum theory
and it also constitutes a central resource in
quantum metrology~\cite{Giovannetti:2004}, quantum communication technologies~\cite{OBrien:2009}, and quantum information processing \cite{Ladd:2010}.
Yet the same principle has been the source of heated discussions since the inception of quantum theory \cite{Schrodinger1935,Born1955,wigner:1962,everett1973,wheeler:1983,Bassi:2000,Zurek:2002,Leggett:2005,Adler:2009,Mermin:2012,Pusey:2012}: the central question of the long-standing debate is about the physical origin of the observed `definiteness' of macroscopic physical objects.
In fact, while it is widely accepted that microscopic systems can live in superposition states, the fact that in a physical apparatus individual measurements always yield single, definite outcomes has so far eluded a comprehensive explanation \cite{Leggett:2008}.
To reconcile the definiteness of measurements with the Schr\"odinger equation, two plausible explanations have been advanced~\cite{Bassi:2013}:
(1) Quantum superposition applies at all scales, even for macroscopic objects, 
and environment-induced decoherence is responsible for the emergence of so-called pointer states, to which the wavefunction is reduced (`collapses') with probabilities determined by Born's rule.
(2) There exists a deeper, underlying theory which gives rise to coherent quantum evolution at the micro scale and yet well-defined trajectories at the macroscopic level, independently of the environment's influence.
This second explanation advocates a `macrorealistic' description of nature as it implies that macroscopic physical objects follow classical trajectories.

In order to put the latter idea of `macrorealism' to the experimental test, Leggett and Garg (LG) derived a set of inequalities bounding the linear combinations of two-time correlation measurements~\cite{Leggett:1985}.
In recent years, violation of LG inequalities has been shown in a wide range of physical systems spanning from superconducting qubits \cite{PalaciosLaloy:2010,Groen:2013}
to photons \cite{White:2011,JinShiXu:2011,Dressel:2011,Suzuki:2012}, nitrogen-vacancy centers in diamond \cite{Wrachtrup:2011}, nuclear spins \cite{Mahesh:2011}, and phosphorus impurities in silicon \cite{Briggs:2012}.
\textcolor{red}{However, these experiments are confined to test superposition states in a simple qubit system, which exhibits Rabi oscillations \textemdash far away from Leggett and Garg's original intention to probe macroscopic quantum superpositions.}

\textcolor{red}{
Performing LG~tests in more complex systems including also mechanical degrees of freedom \textemdash mechanical superposition states are the essential component of most macrorealistic models {\cite{Ghirardi:1986,Pearle:1989,Penrose:1996}} \textemdash constitutes a major challenge:} not only quantum superposition states become very fragile, but also new experimental methods must be developed to realize so-called `ideal negative measurements' in these systems.
Ideal negative measurements \textemdash namely, the ability to measure the physical object yet avoiding any direct interaction with it \textemdash are a prerequisite for any rigorous LG~test, as without it, violations can simply be attributed to an unwitting invasiveness on behalf of the experimenter, rather than to the absence of a realistic description~\cite{Wilde:2012}.
Despite their importance, a rigorous implementation of this type of measurement has been demonstrated in just one of the many LG~tests reported in the literature \cite{Briggs:2012}.

In this article, we report on a \textcolor{red}{$\num{6}\,\sigma$ (standard deviation) violation} of \LGI{} for a cesium atom performing a so-called `quantum walk', in which the atom is coherently transported along a line in discrete steps in space and time.
We obtain the violation by measuring the correlation between the atom's positions at successive times with measurements of the ideal negative type, which a devout realist would perceive as non-invasive.
Our protocol for ideal negative measurements rests upon a novel atom transport technology consisting of two optical lattice potentials that are fully independent, though perfectly stabilized to each other.
The capability of the new system to state-dependently displace atoms over arbitrary large distances enables us to remove atoms depending on their position state and to realize, thus, a negative position measurement on the unshifted atoms.

\textcolor{red}{
Criteria for the assessment of the degree of macroscopicity of superposition states have long been  discussed in the literature \cite{Leggett:2002,Nimmrichter:2013}.
There is a general agreement that the macroscopicity of a mechanical system increases with heavier masses and larger spatial separations of the superposition states.}
Although the atomic wavefunction of the cesium atom in our experiment spreads, at most, over a distance of $\num{5}$ sites ($\SI{2}{\micro\meter}$), our results set the stage \textcolor{red}{for future experiments testing the LG~inequalities with objects of thousands of proton masses split over macroscopic distances (for a review see \cite{Arndt:2014}).}
Furthermore, we remark that this work extends the experimental study of LG~violations to quantum transport systems \cite{Lambert:2010} with dynamics far richer than those of the hitherto-considered qubit systems.

\section{Quantum transport}
Introduced by Richard Feynman to model the one-dimensional motion of a spin-1/2 particle \cite{FeynmanBook}, discrete-time quantum walks can be regarded as the archetype of quantum transport experiments.
While quantum walks share many similarities with classical random walks,
the behavior of these two transport paradigms is strongly different.

In a `classical' random walk scenario, a particle moves in discrete steps, either leftward or rightward, with the direction determined by the result of a coin toss.
After iterating the sequence of coin toss and subsequent displacement $n$ times, one finds the binomial distribution {\arraycolsep=0pt\def\arraystretch{0.6}$\big({\begin{array}{c}{\mbox{\footnotesize$n$}}\\{\mbox{\footnotesize$x$}}\end{array}}\big)\hspace{-0.5pt}/2^n$} describing the motion of the particle by simply enumerating the trajectories terminating in position $x$.
The Brownian motion of colloidal particles suspended in a liquid is a well-known example of this type of diffusive classical transport.

\begin{figure}
	\centering
	\includegraphics[width=\columnwidth]{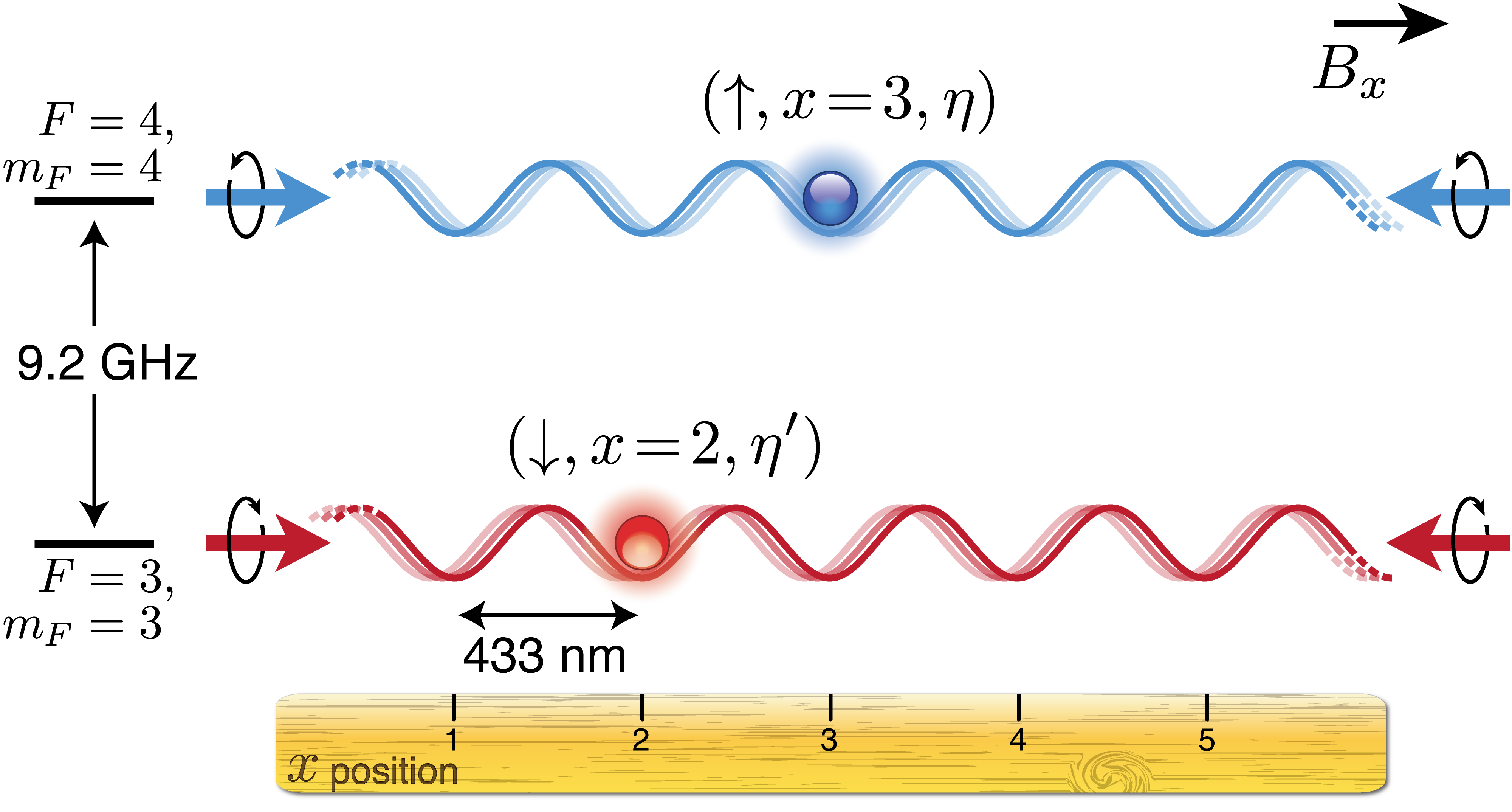}
	\caption{\label{fig:Figure1}\textbf{Transport of single Cs atoms in state-dependent periodic potentials.} Two independent optical lattices originate from standing waves \textcolor{red}{of opposite circular polarization, but identical wavelength $\lambda=\SI{866}{\nano\meter}$}. Depending on the internal state, $\uparrow$ or $\downarrow$, atoms experience one or the other lattice potential. An optoelectronic servo-lock loop allows the position of each lattice to be arbitrarily controlled. The atom's position is retrieved with single site resolution by fluorescence imaging. The parameter $\eta$ accounts for other degrees of freedom, such as the \textcolor{red}{atom's position perpendicular to the lattice} or, in general, other hidden physical aspects. \textcolor{red}{The quantization axis is defined by the small bias magnetic field $B_x$, which is chosen along the two optical lattices. $F$ and $m_F$ denote, respectively, the total angular momentum and its projection along the quantization axis for both internal hyperfine states.}
	}
\end{figure}

\begin{figure*}
	\centering
		\includegraphics[width=\textwidth]{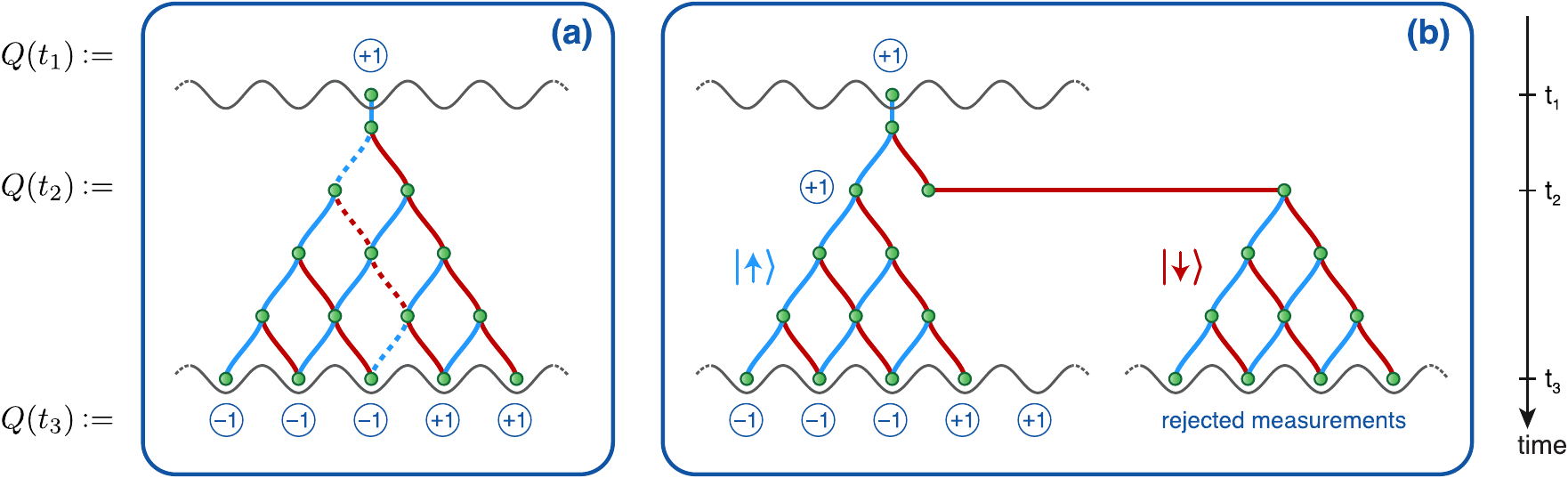}
	\caption{\label{fig:Figure2}\textbf{Ideal negative measurements test the non-classicality of quantum walks.} (a) Schematic representation of a four-step quantum walk containing 16 possible trajectories, which according to quantum mechanics the Cs atom simultaneously follows. Conversely, upholders of realism believe that in each experiment the atom follows a definite trajectory connecting the initial and final point, \emph{e.g.}, the dashed line shown in figure. The outcomes $\pm1$ of $Q(t_i)$ measurements are indicated with circles, where $Q(t_1)$ is identified with the initial state preparation, whereas $Q(t_2)$ and $Q(t_3)$ are related to position measurements. For instance, measurements at times $t_1$ and $t_3$ yields the correlation function $\langle{}Q(t_3)Q(t_1)\rangle$. (b) To measure the correlation function $\langle{}Q(t_3)Q(t_2)\rangle$, we use at time $t_2$ an ideal negative measurement scheme, which ensures the non-invasiveness of $Q(t_2)$: On condition that only atoms in $\downarrow$ state are transported at $t_2$ far away to the right, atoms in $\uparrow$ continue afterwards their walk undisturbed. In case $Q(t_2)$ measurement has not removed the atom, measuring at $t_3$ the atom's position yields $Q(t_3)$ conditioned to the state $(\uparrow,x=-1)$ at $t_2$. Likewise, we obtain $Q(t_3)$ conditioned to $(\downarrow,x=+1)$ by transporting at $t_2$ the atom in $\downarrow$ far away to the left (not shown in the figure).}
\end{figure*}

A different scenario \textemdash which we call `quantum' in the light of the anticipated violation of the \LGI{} \textemdash is instead realized by a cesium atom, which undergoes quantum diffusion in a one-dimensional optical lattice potential.
Rather than tossing a real coin, a microwave `coin' pulse $\mathcal{C}$ is used to put the particle into an equal superposition of two internal hyperfine states \textcolor{red}{of the electronic ground state}, $\ket{F=4,m_F=4}$ and $\ket{F=3,m_F=3}$, which we label for the sake of convenience as $\uparrow$ and $\downarrow$, respectively.
While a quantum physicists would describe $\mathcal{C}$ as a $\pi/2$ rotation of a pseudo spin-1/2 system, a devout realist
would interpret $\mathcal{C}$ as a stochastic process that prepares the atom in one of the two internal states with equal probability\textemdash just like the coin toss.
A state-dependent shift operation $\mathcal{S}$ subsequently moves the atom by one site rightward or leftward depending on the internal state.
As a result of this operation, an atom which is in $\uparrow$ state moves from $x$ to $x-1$, while an atom in $\downarrow$ state moves to $x+1$ instead.
\textcolor{red}
{
The different sensitivity (AC polarizability) of the $\uparrow$ and $\downarrow$ states to left- and right-handed polarized light can be exploited for controlling the atom's position with state-dependent optical potentials, each of which acts on either one of the two internal states (see also Appendix~\ref{appdix:apparatus}) \cite{Mandel:2003}.
As illustrated in Figure~\ref{fig:Figure1}, this idea permits to realize the shift $\mathcal{S}$ by means of two state-dependent optical lattices, whose position is independently controlled with subnanometer precision.
Hence,} the alternation of $\mathcal{C}$ and $\mathcal{S}$ operations realizes a one-dimensional discrete-time quantum walk. 

\begin{figure}[t]
	\centering
		\includegraphics[width=\columnwidth]{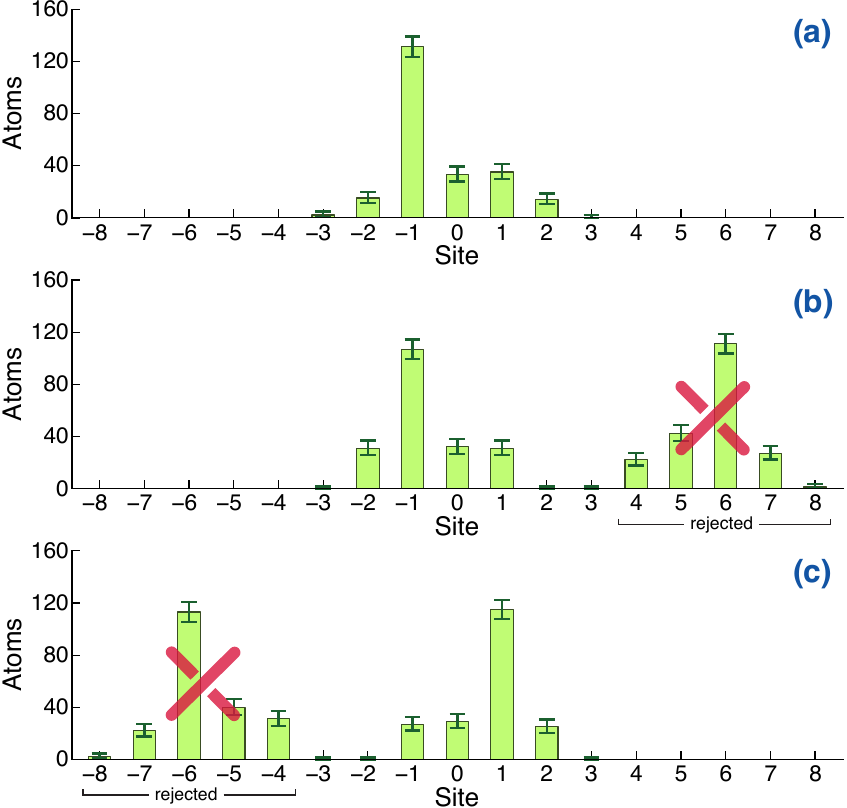}
	\caption{\label{fig:Figure3}\textbf{Violation of Leggett-Garg inequality probing a four-step quantum walk.} The spatial distribution of single atoms is reconstructed by measuring their positions at time $t_3$: \textbf{(a)} If we do not observe which trajectory the atom has taken at $t_2$, the distribution exhibits a pronounced peak on the left hand side. However, when we conclude from an ideal negative result whether the atom at time $t_2$ was in \textbf{(b)} $x=-1$ or \textbf{(c)} $x=1$, we obtain two distributions which resemble the mirror image of one other.
The events in which the atom's position has been affected by $Q(t_2)$ measurement are recognized through the larger displacement and, thus, rejected.
Because the overall number of probed atoms, 404, is the same in (b) and (c),
the retained events can be added together to produce the position distribution at $t_3$ conditioned on having measured the position at $t_2$.
The sum distribution (not shown) is symmetric and differs strongly from the asymmetric distribution in (a).
The vertical error bars represent $\SI{68}{\percent}$ Clopper-Pearson confidence intervals.}
\end{figure}

As revealed in our first im\-ple\-men\-ta\-tion of quantum walks~\cite{Karski:2009}, as well as in several other implementations using diverse physical systems~\cite{Schmitz:2009,Zahringer:2010,Broome:2010,Schreiber:2010,Sansoni:2012}, the spatial probability distribution of the quantum walk expands linearly with the number of steps $n$, in stark contrast to the $\sqrt{n}$ behavior of the classical random walk.
Furthermore, prominent peaks are visible on either one or both sides of the distribution, depending on the initial internal state.

Quantum mechanics gives precise account of these phenomena in terms of interference of all trajectories that the particle is allowed to follow while moving from the initial to the final point.
The agreement with experimental observations, in the spirit of Francis Bacon's inductive thinking,
serves as an important piece of validation of quantum theory itself. However, according to Karl Popper's point of view, one must acknowledge that the remarkable fit between observations and quantum theory does not itself constitute a `falsification' of the `other' hypothesis \textemdash
that an underlying probability distribution could conceivably describe, at all times, the position and the spin of the atom as elements of objective reality.

\section{Leggett-Garg inequality} Here is where the \LGI{} becomes important, as it subjects the idea of realism to a rigorous, objective test by looking for violation of
\begin{equation}
	\label{eq:LGI}
	K\hspace{-2pt} =\hspace{-1pt} \langle Q(t_2) Q(t_1) \rangle\hspace{-1pt} +\hspace{-1pt} \langle Q(t_3) Q(t_2) \rangle \hspace{-1pt}- \hspace{-1pt}\langle Q(t_3) Q(t_1) \rangle\hspace{-1pt} \leq \hspace{-1pt}1,
\end{equation}
where $Q(t_i)$ are real values with $|Q(t_i)|\leq 1$ assigned to the outcomes of a measurement
performed at time $t_i$ with $t_i<t_{i+1}$,
and where $\langle\ldots\rangle$ denotes the average over many repetitions of the experiment.
The derivation of this inequality essentially rests on two assumptions \cite{Emary:2013}: (A1)~realism, as above; and (A2) non-invasive measurability, which asserts the possibility to measure the system without affecting its future evolution.
Both these assumptions are implicit in a realistic view of nature~\cite{Leggett:1985}; but of course, quantum mechanics holds to neither~\cite{vonNeumann:1955,Lueders:1950}.
However, to be a valid test of the LG~inequality, it is sufficient to persuade whom already believes in (A1) that the measurement scheme used in the experiment complies with (A2).
Otherwise, violations of Eq.~(\ref{eq:LGI}) may be attributed to a trivial invasivity of the measurement~\cite{Wilde:2012}.
To ensure this, Leggett and Garg put forward the concept of `ideal negative measurements' \cite{Leggett:1985}, which are well illustrated by the following example:
Imagine that a physical object, like the atom, can be found in only two positions, $x=\pm1$, and that we check the presence of the object at $x=+1$ without looking at $x=-1$.
From the point of view of a realist, the absence of the object at $x=+1$ necessarily implies that $x=-1$ without ever having influenced the object during the measurement.
By repeating this measurement many times, probing the object either at $x=+1$ or $x=-1$ and discarding all measurements that directly reveal the object, we can thus measure correlations functions like $ \langle Q(t_3) Q(t_2) \rangle$ without having ever meddled with the object itself at time $t_2$.
Hence, any violation of Eq.~(\ref{eq:LGI}) that arises from ideal negative measurements must imply a violation of the realist principles (A1) or (A2) \textemdash or both.

\section{Quantum walks falsify classical trajectories}
We base our experiment on a four-step quantum walk probed at times $t_1=0$, $t_2=1$, and $t_3=4$ steps, as displayed in the panels of Figure~\ref{fig:Figure2}, \textcolor{red}{where each step lasts around $\SI{26}{\micro\second}$.}
The three different measurements are defined as follows.
We equate the first measurement $Q(t_1)$ with the state preparation in $(\uparrow,x=0)$:
fluorescence imaging first determines the initial position of the atom with single site resolution \cite{KarskiImaging:2009}, while sideband cooling slows the atom's motion to the lowest longitudinal vibrational state and concurrently polarizes the atom in $\uparrow$ state \cite{Belmechri:2013}.
The translational symmetry of the optical standing wave allows us to safely label the initial position with $x=0$.
We designate $Q(t_1)=1$.
At time $t_2$, we measure the atoms' state, which is restricted to two possibilities, either $(\uparrow,x=-1)$ or $(\downarrow,x=+1)$, and 
we assign to this measurement the value $Q(t_2)=1$ \textcolor{red}{independently of the atom's internal state or position}.
The assignment of $Q(t_2)$ to a constant value is, in fact, one of the legitimate choices that are consistent with the condition $|Q(t_i)|\le 1$ in the derivation of LG~inequalities~\cite{Emary:2013}.
Finally, $Q(t_3)$ measures the atom's position at the end of the walk and returns the value $-1$ for $x\leq{}0$ and value $+1$ for $x>0$.
According to quantum mechanics, with this definition of $Q(t_i)$ we expect a violation of the \LGI{} yielding $K=\num{1.5}$ \textcolor{red}{(see Appendix \ref{appdix:theory})}.

\textcolor{red}{
Quantum mechanics also shows that other designations of $Q(t_2)$ are possible to produce a violation of Eq.~\ref{eq:LGI}, for instance, by assigning the measurement outcome $(\downarrow,x=+1)$ to $1$ and $(\uparrow,x=-1)$ to a certain value $\xi$ with $|\xi|\le1$.
While previous experiments \cite{PalaciosLaloy:2010,Groen:2013,White:2011,JinShiXu:2011,Dressel:2011,Suzuki:2012,Wrachtrup:2011,Mahesh:2011,Briggs:2012} have adopted a dichotomic designations of $Q(t_2)$ (analogous to set here $\xi=-1$), we have intentionally dropped such an extra condition to permit larger violations of the LG~inequality as fewer constraints are imposed (cf.\ Eqs.~\ref{eq:nodichotomic} and \ref{eq:dichotomic} in Appendix~\ref{appdix:theory}).
Such a constant designation especially reveals that the essential requisite to violate Eq.~\ref{eq:LGI} is that the particle is measured at $t_2$, even though the result of the measurement itself is then discarded.
}

Because the measurement $Q(t_1)$ is a state preparation, and because we are not concerned about the atom's evolution after time $t_3$, only the measurement $Q(t_2)$ must be performed non-invasively.
Since we are not allowed to directly image the atom at time $t_2$ because it would be invasive, we adopt an ideal negative measurement strategy that hinges on state-selective removal of atoms.
This measurement scheme draws direct inspiration from the experimental realization of interaction-free measurements of the state of single photons \cite{Kwiat:1995}.
The measurement scheme, which is illustrated in Figure~\figref{fig:Figure2}{b}, proceeds as follows:
\textcolor{red}{if we want to non-invasively detect the atom's presence, say, in $x=-1$}, we remove the atoms in the state $(\downarrow,x=+1)$ by transporting them far to the right, whereas we leave the atoms in the state $(\uparrow,x=-1)$ untouched.
Providing this shift (set here to 5 sites) is larger than the distance covered by the atom between $t_2$ and $t_3$, the atom's position at the later time $t_3$ allows us to unequivocally mark the shifted atoms (which though remain trapped in the lattice potential) as effectively removed with confidence better than $99\%$.
Hence, the state-selective removal of atoms provides information about the atom's position at time $t_2$ and, at the same time, postselects those measurements that are carried out non-invasively.

\textcolor{red}{In the experiment, state-selective removal of atoms requires the ability to shift one single spin species at a time over arbitrary distances.
However, previous implementations of state-dependent transport have so far only demonstrated the concurrent shift of both spin species instead of an individual one  \cite{Mandel:2003,Karski:2009,Genske:2013}.
Moreover, the largest displacement attained heretofore with a single transport operation amounts to about one lattice site \cite{Mandel:2003}.
We overcome these limitations by employing a new atom transport technology, which relies on two spatially overlapped, yet fully independent optical lattices.
In the new implementation, the two optical standing waves that create the lattice potentials (see also Figure~\ref{fig:Figure1}) originate from independent laser beams with opposite circular polarizations, whose phase and frequency can individually be controlled with the aid of acousto-optic modulators. 
%
%
Two optical phase-lock loops are employed to stabilize the position of both periodic lattices against a common third reference laser beam.	
We thereby achieve a stability of the relative position between the two lattices on the level of $\SI{100}{\pico\meter}$ to be compared with the $\SI{20}{\nano\meter}$ localization of the atoms along the lattice direction.
The complete independence of the two standing waves allows us to arbitrarily control the position of each lattice by varying the phase of the corresponding laser beams.
}
The intensity of each laser beam is actively stabilized to better than $\SI{0.1}{\percent}$ RMS noise.
%

In order to measure the LG correlation function, we note that with our assignment of $Q(t_i)$, the correlation function $K_{12}\equiv \langle Q(t_2) Q(t_1) \rangle$ is trivially equal to one.  Furthermore, we have $K_{13}\equiv \langle Q(t_3) Q(t_1) \rangle = \langle Q(t_3) \rangle$, which quantifies the asymmetry of the final position distribution. Figure~\figref{fig:Figure3}{a} shows the measured probability distribution of a four-step quantum walk with fair coin toss ($\theta = \pi/2$). The distribution is characterized by a pronounced skew to the left, which translates into a non-zero value $K_{13}=\num{-0.57\pm0.05}$.
Although this asymmetry itself is often interpreted as a hallmark of `quantumness' \cite{Karski:2009,Schmitz:2009}, we would rather eschew similar premature conclusions here.
Using the law of total probability under assumptions (A1) and (A2), the final correlation function may be obtained as
\begin{equation}
  \label{eq:K23}
  K_{23} 
  =  
  \sum_{x=\pm1} P(t_2;x) \hspace{1pt} \langle Q(t_3) \rangle_{x}
,
\end{equation}
where $P(t_2;x)$ is the probability of finding the atom in $x$ at $t_2$, and $\langle\ldots\rangle_{x}$ is the average over the distribution conditioned on a negative detection of the atom in $x$ at $t_2$.
Hence, we perform two separate experiments to measure $K_{23}$, one for each term of the sum in Eq.~(\ref{eq:K23}), as shown in Figure~\figref{fig:Figure3}{c}.
After rejecting all measurements during which atoms have provably been perturbed, we find $P(t_2;x=-1)=\SI{0.506\pm 0.026}{}$ and $P(t_2;x=+1)=\SI{0.494\pm 0.026}{}$.
Averaging $Q(t_3)$ with the two conditioned distributions yields a value $K_{23}=\num{-0.14\pm0.05}$ close to zero. Taken together, the three correlation functions yield $K=\num{1.435\pm0.074}>1$, which violates the \LGI{} by about $6\,\sigma$. The uncertainty is estimated to be purely statistical (see Appendix~\ref{appdix:staterrors}).

\begin{figure}[b]
	\centering
		\includegraphics[width=\columnwidth]{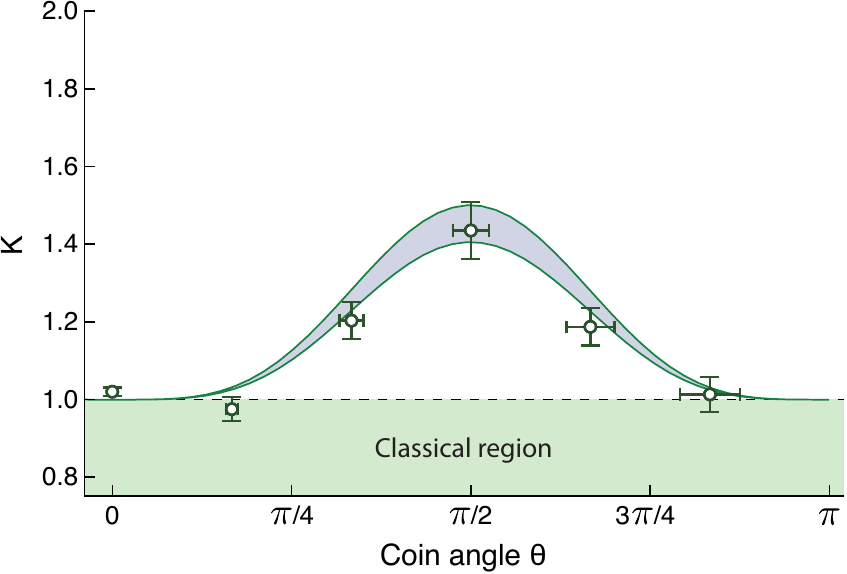}
	\caption{\label{fig:Figure4}\textbf{Leggett-Garg correlation measurement witnessing the degree of quantumness.} 
Maximum violation occurs for a fair coin ($\theta=\pi/2$), while no violation occurs for classical transport at $\theta=0$ and $\theta=\pi$.
The solid lines is the theoretical prediction based on quantum mechanics of the LG correlation function $K$ for a decoherence free quantum walk (upper curve) and for a quantum walk with $\SI{10}{\percent}$ decoherence per step (lower curve). The vertical error bars represent $1\,\sigma$ uncertainty, while the horizontal error bars denotes a systematic uncertainty on the coin angle.}
\end{figure}

\section{Quantum witness}
Besides the fundamental interest, LG~inequalities also find application in quantifying the degree of `quantumness' of a system.
This requires, however, that we abandon the standpoint of realists and, from now on, embrace quantum mechanics instead.
Intuitively, the LG correlation function $K$ may serve as an indicator, say a witness, of the amount of superposition involved in the system's dynamics.
This idea of `quantum witnesses' has recently been proposed as a method to discern quantum signatures in systems like biological organisms \cite{Che-Ming:2012}.

Owing to our particular definition of $Q(t_2)$, which is constantly mapped to $1$, we prove a direct connection \textcolor{red}{(see Appendix~\ref{app:qntwitness})} between LG~inequalities and quantum witness formalism by identifying $W \equiv |K-1|$ with the first quantum witness introduced in Ref.~(\citenum{Che-Ming:2012}).
The deviation of $W$ from zero indicates the degree of `quantumness' in the system's dynamics.

We provide demonstration of the quantum witness $W$ in the four-step quantum walk by testing different types of coins, which differ in the probability of tails $p=\cos^2(\theta/2)$ and heads $q=1-p$.
For instance, $p=q=\num{1/2}$ corresponds to the fair coin's situation, which has hitherto been considered.
As displayed in Figure~\ref{fig:Figure4}, we measure the LG correlation function $K$ for different values of the coin angle $\theta$, which is tuned by setting the duration of the coin's microwave pulse. The violation is maximal for $\theta=\pi/2$ (fair coin), \textcolor{red}{when the coin maximally splits the walker's state at each step in an equal superposition of states.} Instead, the violation vanishes for $\theta = 0$ and $\theta = \pi$, when the walk reduces to classic transport with no superposition involved.

\section{Interpretation and discussion}
The reported violation of the \LGI{} proves that the concept of a well-defined, classical trajectory is incompatible with the results obtained in a quantum walk experiment.
Yet, the concept of well-defined trajectories in position space can, in part, still be rescued providing one renounces locality.
An example is provided by Bohmian mechanics, whose predictions are shown equivalent to those of non-relativistic quantum mechanics~\cite{bohmian:2009}.
In this interpretation of quantum theory, physical objects follow precise trajectories, which are guided by the universe's pilot wavefunction, that is, by a physical entity constituting a non-local hidden variable.
It is therefore clear that Bohmian mechanics is not in contradiction with our findings since, from that point of view, assumption (A2) is not fulfilled.

\textcolor{red}{
Recently, a minimal macrorealistic extension of (non-relativistic) quantum mechanics has been put forward under general assumptions \cite{Nimmrichter:2013}, which proposes a universal objective measure of macroscopicity accounting for both the mass and spatial separation of the superposition states.
Within this model, we estimate a measure of macroscopicity for our experiment (see Appendix~\ref{app:macmeasure}) that lies in the range of typical cold atom experiments \cite{Arndt:2014} \textemdash whether those be performed with thermal atoms or with a Bose-Einstein condensate.
We remark, moreover, that the macroscopicity of our experiment is, coincidentally, on the same order of magnitude of experiments testing superpositions of macroscopic persisting currents \cite{PalaciosLaloy:2010,Arndt:2014,Korsbakken:2010}.
In spite of the yet microscopic nature of the present LG tests, our result gives a conceptual demonstration that non-invasive measurement techniques can be applied to test the LG~inequality, \emph{e.g.}, in double-slit experiments with genuinely massive particles by alternatively blocking at time $t_2$ either one of the two slits.
%
}

Unlike the test of Bell inequalities, where a loophole-free violation seems in reach \cite{Merali:2011}, LG~experiments remain susceptible to the so-called clumsiness loophole \textemdash even employing negative measurements.
This loophole refers to the impossibility on behalf of the experimenter to exclude an invasivity of the measurements.
Hence, it is appropriate to comment on the three main instances which can hinder the fulfillment of (A2) in our experimental set-up.
(1) In the measurement of $Q(t_2)$, the state-dependent shift could cause motional excitations to the unshifted atoms. To avoid this problem, we deliberately set the shift duration to a time of $\SI{200}{\micro\second}$, which is much longer than than the period of the longitudinal motion of $\SI{\approx10}{\micro\second}$.
We measured the fraction of atoms that are left in the ground state by the shift process for both shifted and unshifted internal states~\cite{Belmechri:2013}. In both cases, we obtained a fraction $>\SI{99}{\percent}$, which is consistent with the precision of the initial preparation, thus confirming that no excitation is produced.
\textcolor{red}{
The concept of venality, which has been introduced in Ref.~\citenum{Briggs:2012} to account for non-ideal negative measurements, can be applied to this effect as well. The analysis in Appendix~\ref{appdix:venality}, however, shows that the upper limit imposed on $K$ is only slightly changed.
}
(2) The duration of measurement $Q(t_2)$ is comparable to the spin coherence time. In principle, an equal delay time should also be included in the sequence when no measurement is performed at $t_2$. Even doing so, we verified using a different experimental sequence, a Ramsey interferometer instead of a quantum walk, that a violation of the \LGI{} is still produced.
(3) At time $t_1$, the motion of the atom in the transverse direction is prepared according to a Boltzmann-like distribution, which extends over the first hundred motional states.
A statistical mixture is not a problem \emph{per se}, providing the statistical properties are maintained constant.
A realist, though, could raise the objection that the experiment `knows' which correlation term, either $K_{13}$ or $K_{23}$, is being measured and exploits this information to  prepare the transverse motion \emph{ad hoc} in a way to counterfeit the violation of the \LGI{} \textcolor{red}{(cf.~the hypothesis of so-called induction discussed by Leggett in Ref.~\citenum{Leggett:2008}).}
More generally, the same argument can also be invoked in case of any hidden variable $\eta$, which, from an epistemological point of view, is tantamount to the transverse motion of the atoms.
Eventually, to blunt this criticism, one could base the choice which correlation term to measure upon random events that are uncorrelated from the initial preparation \cite{Scheidl:2010,Gallicchio:2014}.

There is one further aspect of this LG~test that must be emphasized, namely that we test single, individual copies of the system by probing one Cesium atom at a time.
Prior experiments in NMR systems~\cite{Mahesh:2011,Briggs:2012} took an alternative approach by substituting individual measurements with measurements on a large ensemble of identical systems instead.
Our approach \emph{a priori} eliminates the need for the extra assumption that multiple copies of the system \textemdash even when positioned in near proximity \textemdash do not interact with each other.
However plausible this hypothesis is in NMR systems, ignoring it would allow a realist to argue that the several copies of the system have interacted with each other \textemdash in particular with those copies that have been invasively measured, thus invalidating hypothesis (A2).
In addition, employing ensembles instead of individual systems can lead to controversial interpretations, as is illustrated by the following examples. A wave-like analogue of quantum walks based on coherent electromagnetic waves (\emph{e.g.}\ a laser beam \cite{Schreiber:2010}) is expected to produce a violation of the \LGI{} similar to the one obtained with individual photons. In a similar way, even acoustic or surface waves could be used to measure a violation. However, it is certainly debatable whether an experiment hinging on Maxwell equations or mechanical waves can indeed rule out realism. In fact, to reach this conclusion, a realist should be first persuaded that light is composed of photons and waves of phonons.

In conclusion, our experiment gives a rigorous, quantitative demonstration of the non-classicality of a massive-particle quantum walk.
The experiment also sets the basis for a test of LG~inequality probing the positional degree of freedom over macroscopic distances.
The interaction-free detection method of the atom's position can well be adapted to other systems like matter wave interferometers with large spatial splitting \cite{Alberti:2009,Muentinga:2013,Dickerson:2013}.
The ten-dimensional Hilbert space (5 lattice sites with 2 internal states each) of this LG~test constitutes a significant advance beyond
the simple two-level system, which has been so far investigated.
Moreover, the multidimensionality of the Hilbert space~\cite{Budroni:2014} can be used in the future to approach the algebraic limit of the correlation function $K$, which is equal to $\num{3}$.
Finally, we should remark the illustrative value of this violation of the \LGI{}, which puts the particle's trajectories in position space at center stage.

\appendix

\section{EXPERIMENTAL APPARATUS}
\label{appdix:apparatus}
Each experimental sequence starts with, on average, $\num{1.2}$ atoms sitting at sufficiently separated lattice sites.
Atoms are cooled to the longitudinal ground state using first molasses cooling and then microwave sideband cooling \cite{Belmechri:2013}, \textcolor{red}{while they are thermally distributed in the direction transverse to the lattice with a temperature of $\approx \SI{10}{\micro\kelvin}$.}
Optical pumping initializes $>\SI{99}{\percent}$ of the atoms in the $\uparrow$ state.
The duration of coin pulses, which are resonant with the hyperfine splitting of $\SI{9.2}{\giga\hertz}$, determines the value of the coin angle $\theta$, with the fair coin pulse lasting $\SI{4.5}{\micro\second}$ (calibrated using Rabi oscillations).
The wavelength $\lambda$ of the optical lattice and the two Zeeman hyperfine states are chosen such that the $\uparrow$ state experiences an \textcolor{red}{optical dipole potential originating only from right-handed circularly polarized photons, while $\downarrow$ state experiences a potential produced by both left- and right-handed circularly polarized photons} with relative weights of $\num{7/8}$ and $\num{1/8}$, respectively.
The lattice depth of $\approx\SI{80}{\micro\kelvin}$
precludes tunneling between different sites.
To implement the state-dependent shift, the two standing waves are displaced by one site with respect to each other with a linear ramp lasting $\SI{21}{\micro\second}$, which leaves $>\SI{99}{\percent}$ of the atoms in the motional ground state, measured with sideband spectroscopy.
During the shift, the potential depth experienced by the $\uparrow$ state remains constant, while the one experienced by the $\downarrow$ state is modulated, with a minimum depth of $3/4$ in relative units \cite{Belmechri:2013}.
At the beginning and the end of each sequence, fluorescence imaging determines the position of individual atoms with a measured reliability of $\SI{98}{\percent}$, while $\SI{2}{\percent}$ of the atoms are erroneously attributed to the adjacent site.

\section{DECOHERENCE ANALYSIS}
A four-step quantum walk lasts around $\SI{100}{\micro\second}$ without including the duration of $Q(t_2)$. This time should be compared with the spin relaxation time ($T_1$) and the spin coherence time ($T_2$). In our system, we measure $T_1=\SI{107\pm5}{\milli\second}$, which is due to Raman scattering of photons from the optical lattice. The $T_2$ time is mainly limited by inhomogeneous dephasing due to magnetic field fluctuations and to both scalar and vectorial differential light shifts. Defining $T_2$ as the duration of the Ramsey sequence whose interference contrast is reduced to $\SI{50}{\percent}$, we measure $T_2=\SI{229\pm16}{\micro\second}$.
We also fit a density matrix description of decohered quantum walks to the measured position distributions \cite{alberti:2014}. Using the amount of spin decoherence per step as the only free fit parameter, we obtain that spin coherence decreases after each step by $\SI{6}{\percent}$ for the fair coin and by $\lesssim\SI{10}{\percent}$ for the other points in Fig.~\ref{fig:Figure3}, with the reduced chi-squared being $\lesssim\num{1}$.

\section{STATISTICAL ERRORS}\label{appdix:staterrors}
In this work, the confidence intervals of the correlation measurements represent $1\,\sigma$ statistical uncertainty, which has been computed by fitting a Gaussian profile to the bootstrapped distribution (\emph{i.e.}\ the distribution obtained by resampling with replacement). Independently from  bootstrapping, we also computed the statistical uncertainties using Monte Carlo resampling, where the statistical errors of position distributions are estimated with binomial statistics (Clopper-Pearson method). The two estimation methods lead to consistent results. For instance, for the fair coin, we obtain $K=\num{1.435\pm0.068}$ with Monte Carlo and $K=\num{1.435\pm0.074}$ with bootstrapping. While Monte Carlo analysis requires invariant statistical properties to be valid, bootstrapping analysis remains valid also in the presence of slow drifts of experimental parameters.
The close agreement between the two statistical analyses indicates that each correlation measurement of $K$ (lasting about $\SI{120}{\minute}$) is performed under constant experimental conditions.

\section{SYSTEMATIC ERRORS}
\textcolor{red}{Systematic errors \textemdash that is, deviations from the ideal quantum walk evolution \textemdash do not invalidate the result of a LG~test provided that the experiment is performed under constant experimental conditions and that hypothesis (A2) is not contradicted.
Nevertheless, we shortly comment on the} three main mechanisms that bring about systematic fluctuations:
(1) Imperfect initialization prepares $<\SI{1}{\percent}$ of the atoms in the wrong internal state. However, to derive the \LGI{}, a statistical mixture defining the initial state is perfectly admissible.
(2) Imperfect reconstruction of the atom's position can be accounted for in terms of a noisy measurement apparatus.
(3) Spontaneous flips of the internal state can be accounted for in terms of an additional stochastic process, which also contributes to determine the system's evolution.
We estimate that each of these three mechanisms actually affects the position distribution by $<\SI{1}{\percent}$, that is less than the statistical uncertainty.

\section{VENALITY}\label{appdix:venality}
\textcolor{red}
{
Knee \emph{et al.}\ has introduced in Ref.~\citenum{Briggs:2012} the concept of venality $\zeta$ to quantify how often a non-ideal negative measurement, \emph{i.e.}\ a measurement that could potentially violate (A2), has been performed.
In our experiment, it occurs with a relative frequency of $\SI{1}{\percent}$ (estimated as the upper limit) that motional excitations of the unshifted atoms are produced during the measurement of $Q(t_2)$. In addition, spontaneous flips of the internal state happening during the $\SI{200}{\micro\second}$-long $Q(t_2)$ measurement could also invalidate hypothesis (A2). This second process, however, occurs with an even smaller relative frequency of $\approx\SI{0.2}{\percent}$. Hence, we quantify the relative frequency of non-ideal negative measurements with $\zeta=\SI{1}{\percent}$.}

\textcolor{red}{
Along the lines of Ref.~\citenum{Briggs:2012}, the correlation function $K$ measured in our experiment can be decomposed as $K=1+(1-\zeta)K_{23}^\text{ideal}+\zeta K_{23}^\text{corrupt}-K_{13}$, where $K_{23}^\text{ideal}$ and $K_{23}^\text{corrupt}$ denote the correlation function $\langle Q(t_3) Q(t_2) \rangle$ which has been measured with an ideal negative measurement $Q(t_2)$ and with a corrupted one, respectively. Taking into account the venality $\zeta$, the Leggett-Garg inequality, which is derived from (A1) and (A2), reads $K\le1+\zeta(K_{23}^\text{corrupt}-K_{13})$. From this we obtain a new upper bound for $K\le1+2\zeta=\num{1.02}$, which is only slightly displaced from the ideal case of $1$.
}\vspace*{-0.1cm}

\section{QUANTUM MECHANICAL PREDICTION}\label{appdix:theory}
\textcolor{red}{A quantum mechanical calculation shows that, among the possible designations of $Q(t_i)$, the maximal violation of LG~inequality is obtained by associating the measurements' results to the extremal values in the permitted range, that is, either $+1$ or $-1$.
Other designations, \emph{e.g.}\ $Q(t_3)=x/2$, would lead to a smaller upper bound for $K$.}

\textcolor{red}{With our prescription of $Q(t_i)$, we find for a four-step quantum walk an analytic expression of $K$ as a function of the coin angle $\theta$,}
\begin{equation}
	\label{eq:nodichotomic}
  K = 
  \frac{1}{16} \left[
    19-4 \cos (2 \theta)+\cos (4 \theta)
  \right],
\end{equation}
which is the curve plotted as the upper line in Fig.~\ref{fig:Figure4}.
%
Alternatively, with a dichotomic assignment of $Q(t_2)$ equal to $-1$ for $(\uparrow,x=-1)$ and to $+1$ for $(\downarrow,x=+1)$, we obtain
\begin{equation}
		\label{eq:dichotomic}
  K = \frac{1}{32} 
  \left[
    33 -4 \cos (\theta)-4 \cos (2 \theta)+4 \cos (3 \theta)+3 \cos (4 \theta)
  \right]
  ,
\end{equation}
which reaches the maximum value of approximately $1.31$, in contrast to $\num{1.5}$ corresponding to Eq.~(\ref{eq:nodichotomic}).
\vspace*{-0.1cm}

\section{QUANTUM WITNESS}\label{app:qntwitness}
The assignment $Q(t_2)=1$, together with $Q(t_1) =1$ by preparation, implies that the LG inequality (\ref{eq:LGI}) can be written in general terms as
\begin{equation}
	K  -  1
	=
     \left(
	\sum_{x=\pm1} P(t_2;x) \hspace{1pt} \langle Q(t_3) \rangle_{x}
     \right)	
      - 
	\hspace{-1pt}\langle Q(t_3) \rangle\hspace{-1pt} 
	\leq \hspace{-1pt}0\,  .
\end{equation}
This inequality is but one of a family of inequalities, and $K'\hspace{-3pt} =\hspace{-1pt} \langle Q(t_2) Q(t_1) \rangle\hspace{-1pt} -\hspace{-1pt} \langle Q(t_3) Q(t_2) \rangle \hspace{-1pt}+\hspace{-1pt}\langle Q(t_3) Q(t_1) \rangle\hspace{-1pt} \leq \hspace{-1pt}1$ defines a similar, though independent inequality built from the same correlation terms~\cite{Emary:2013}.  With the choice of $Q(t_i)$ discussed here, we find that $K'-1 = -(K-1)$. Taken together, these two inequalities imply that $W=|K-1| = 0$.  The comparison with Ref.~\citenum{Che-Ming:2012} allows us to identify $W$ as the first quantum witness in that work.

\vspace*{1cm}
\section{MACROSCOPICITY MEASURE}\label{app:macmeasure}
\textcolor{red}{
Nimmrichter \emph{et al.}\ \cite{Nimmrichter:2013} have suggested a universal, objective measure $\mu$ that quantifies the amount of macroscopicity of a mechanical superposition state.
In the proposed model, $\mu$ sets a lower limit for the time (expressed in logarithmic scale) during which an electron \textemdash chosen as the reference particle \textemdash behaves like a ``wave'' delocalized over distances larger than a certain critical classicalization length scale $\ell$, which represents a phenomenological parameter.
The length scale $\ell$ is defined in the model such that quantum superpositions of paths separated by less than $\ell$ preserve their coherence.
We estimate for our experiment $\mu = \log_{10}(T\hspace{2pt}M^2_{\text{Cs}}/m_\text{e}^2) \approx 6.8$ for values of $\ell$ shorter than the maximal separation, $\SI{2}{\micro\meter}$, reached during the 4-step quantum walk.
Here, $M_\text{Cs}$ and $m_\text{e}$ denote the masses of the Cs atom and of an electron, respectively, and $T$ represents the overall duration of the quantum walk.
For values of $\ell$ larger than $\SI{2}{\micro\meter}$, the measure $\mu$ as a function of $\ell$ itself behaves, up to an additive constant, as $-2\log_{10}(\ell/\SI{2}{\micro\meter})$ \cite{Nimmrichter:2013}.
}


\begin{acknowledgments}
We are indebted to Jean-Michel Raimond for insightful discussions during his stay in Bonn and for critical reading of the manuscript.
We thank Ricardo Gomez for his contribution to the experimental apparatus.
We acknowledge financial support from NRW-Nachwuchsforschergruppe ``Quantenkontrolle auf der Nanoskala'', ERC grant DQSIM, EU project SIQS. In addition, AA acknowledges support from the Alexander von Humboldt Foundation, and CR from BCGS program and Studienstiftung des deutschen Volkes.
\end{acknowledgments}

\ifusebibfile

\bibliographystyle{apsrev4-1}
\bibliography{leggett-garg}

\else

\fi

\end{document}